\documentclass[a4paper,fleqn,usenatbib]{mnras}

\usepackage[english]{babel}
\usepackage[utf8x]{inputenc}
\usepackage{amsmath}
\usepackage{geometry}
\usepackage[colorinlistoftodos]{todonotes}

\usepackage{todonotes}
\usepackage{easyReview}

\usepackage{color}
\usepackage[T1]{fontenc}
\usepackage{ae,aecompl}
\usepackage{newtxtext,newtxmath}
\usepackage{graphicx}	
\usepackage{amsmath}	
\usepackage{amssymb}	

\title{Abundance Tomography of Type Iax SN 2011ay with TARDIS}
\author[Barna et al.]{Barnab\'as Barna$^{1}$, Tam\'as Szalai$^{1}$, Markus Kromer$^{2,3}$, Wolfgang E. Kerzendorf$^{4}$, \newauthor J\'ozsef Vink\'o$^{5,1,6}$, Jeffrey M. Silverman$^{6}$, G. H. Marion$^{6}$, J. Craig Wheeler$^{6}$
\\
$^{1}$Department of Optics and Quantum Electronics, University of Szeged, Dom ter 9 Szeged, Hungary\\
$^{2}$Zentrum für Astronomie der Universität Heidelberg, Institut für Theoretische Astrophysik, Philosophenweg 12, D-69120 Heidelberg, Germany\\
$^{3}$Heidelberger Institut für Theoretische Studien, Schloss-Wolfsbrunnenweg 35, D-69118 Heidelberg, Germany\\
$^{4}$European Southern Observatory, Karl-Schwarzschild-Strasse 2, 85748 Garching bei München, Germany\\
$^{5}$Konkoly Observatory of the Hungarian Academy of Sciences, Konkoly Thege Miklós út 15-17., Budapest H-1121, Hungary\\ 
$^{6}$Department of Astronomy, University of Texas at Austin, 1 University Station C1400, Austin, TX 78712-0259, USA\\
}

\begin{document}
\maketitle

\begin{abstract}
We present a detailed spectral analysis of Type Iax SN 2011ay. Our spectra cover epochs between -3 and +19 days with respect to the maximum light in B-band. This time range allows us to employ a so-called abundance tomography technique. The synthetic spectral fitting was made with the 1D Monte Carlo radiative transfer code TARDIS.
In this paper, we describe our method to fit multiple epochs with a self-consistent, stratified atmospheric model. We compare our results to previously published SYN++ models and the predictions of different explosion scenarios. Using a fixed density profile (exponential fit of W7), we find that a uniform abundance model cannot reproduce the spectral features before maximum light because of the emergence of excessively strong Fe lines. In our best-fit TARDIS model, we find an abundance profile that separated into two different regimes: a well-mixed region under 10,000 km s$^{-1}$ and a stratified region with decreasing IGE abundances above 10,000 km s$^{-1}$.
Based on a detailed comparative analysis, our conclusion is that the available pure deflagration models cannot fully explain either the observed properties of SN~2011ay or the results of our TARDIS modeling. Further examinations are necessary to find an adequate explanation for the origin of this object.

\end{abstract}


\section{Introduction} \label{introduction}
A large group of Type Ia supernovae (SN Ia) form a nearly homogeneous class of 'standardizable candles' that originates from thermonuclear explosions of white dwarf stars \citep[WDs; e.g.][]{Hoyle60,Colgate69,Nomoto84}. Nevertheless, with the increasing amount of observational data it became clear that SN Ia are not a homogeneous class of objects, but there are a number of spectroscopically peculiar subgroups. One of these groups are the 2002cx-like objects, whose first identified members were the prototype SN~2002cx \citep{Li03,Branch04,Jha06}, and SN~2005hk \citep{Chornock06,Phillips07}. 
To date, $\sim$50 SNe have been classified as members of the group, which has been declared a separate class of stellar explosions, called SNe Iax, by \citet{Foley13}. Based on their work and references therein, the main observational properties of SNe Iax are low expansion velocities (typically between 5,000 and 8,000 km s$^{-1}$), low values of peak absolute brightness (-14 mag > $M_\rmn{V,peak}$ > -19 mag), a lack of high-velocity features (HVFs) in their early-time spectra, and the presence of permitted Fe\,\begin{small}II\end{small} lines near maximum and even weeks after that. 

Despite the obvious observational similarities, SNe Iax hardly can be taken to be a homogeneous class. 
Conspicuous evidence for this statement is the broad range of peak magnitudes and velocities denoted above. Although, in general, low velocity seems to correlate with low luminosity in the case of SNe Iax, some of them do not follow this trend; for example, both SN~2009ku \citep{Narayan11,Foley13} and SN~2014ck \citep{Tomasella16} were extremely slow (with $v_{\rmn{peak}}$ $\sim$ 3,000 km s$^{-1}$), but relatively high-luminosity events.
There are also open questions concerning the locations of these stellar explosions. Almost all known SNe Iax have been found in late-type host galaxies with recent, strong star formation activity, except SN~2008ge \citep{Foley10,Foley13} and maybe SN~2002bp, SN~2005cc, and iPTF~13an \citep[in these cases, the red colours of the host galaxies indicate old environments, see][]{White15}. 

These similarities and differences among SNe Iax make the question of the possible progenitor systems especially interesting: do their commonalities hint at the same explosion mechanism, or are there ''subclasses'' within the group of SNe Iax?
Looking at the existing theories, the most popular scenario is the pure deflagration of a carbon-oxygen (CO) Chandrasekhar mass ($M_\rmn{Ch}$) WD with a bound remnant (called also ``failed'' deflagration), which is supported by the results of hydrodynamic explosion models \citep{Jordan12b,Kromer13,Fink14} as well as by detailed spectral modeling \citep{Magee16} and binary populations synthesis calculations \citep{Liu15}. A modified version of this theory, in which the progenitor is a hybrid CONe WD, has been also published in several papers to explain the explosion mechanisms of the faintest members of the class \citep[see e.g.][]{Denissenkov15,Kromer15,Bravo16}; however, very recent results indicate that such hybrid WDs become unstable to mixing on a very short timescale, which can affect the final fate of these objects \citep{Brooks17}. 

A deflagration that fails to completely unbind the (progenitor) WD is broadly consistent with the general observed characteristics of SNe Iax and provides a clear explanation to the wide range of kinetic energies and peak brightnesses. 
Nevertheless, there are some open questions that leave room for further investigation. One of these problems is that synthetic light curves from the deflagration models of \citet{Kromer13} and \citet{Fink14} seem to evolve too fast after their maxima, indicating excessively low optical depths, i.e. insufficient ejecta masses.
Nevertheless, it has to be noted that only a small range of initial parameters of deflagration models has been sampled in these studies; therefore, it cannot be excluded that with different initial conditions (ignition geometry, density, composition) larger ejecta masses (for a given $^{56}$Ni mass) are possible in the deflagration scenario.

Another problematic point is the structure of the ejecta. The lack of the secondary peak in the red and near-infrared light curves of SNe Iax as well as the late-time spectroscopic characteristics of some of these objects indicate that there is a strong mixing of elements in their ejecta \citep{Jha06,Phillips07}. One of the main motivations of testing pure deflagration models was that these predict such homogeneous ejecta structures -- instead of any explosion models involving detonation result in a (partially) stratified abundance distribution. At the same time, the pure deflagration picture contradicts the bright SN Iax 2012Z \citep{Stritzinger15}, in which the velocity distribution of intermediate mass elements (IMEs), e.g. of silicon and magnesium, do not indicate significant mixing in the outer ejecta.

In the frame of single-degenerate (SD) explosion scenarios, the formation of a (partially) layered ejecta structure can be explained with a mechanism in which the deflagration turns into detonation, due to a the deflagration-
to-detonation transition \citep[DDT,][]{Khokhlov91a}. In classical DDT models, the WD becomes unbound during the deflagration phase; in a variation, the pulsational delayed detonation (PDD) models, the WD remains bound at the end of the deflagration phase, then undergoes a pulsation followed by a delayed detonation \citep{Ivanova74,Khokhlov91b,Hoeflich95}. Based on other assumptions,  detonation can happen via a sudden energy release in a confined fluid volume; this group of models includes gravitationally confined detonations \citep[GCD, see e.g.][]{Plewa04,Jordan08}, ''pulsationally assisted'' GCD models \citep[PGCD,][]{Jordan12a}, and pulsating reverse detonation (PRD) models \citep{Bravo06,Bravo09a,Bravo09b}.
From all of these delayed detonation scenarios, only PDD models have been used for direct comparison with the data of an SN Iax \citep[SN 2012Z,][]{Stritzinger15}. While in this single case, PDD seems to be a viable explanation, these models and others mentioned above need to be compared to a larger sample of Iax explosions.

Assuming binary systems where WDs accrete He-rich material from their companions widens the field of possible SD explosion scenarios of SNe Iax even more. Such progenitor systems have been proposed first by \citet{Foley13}, who suggested the detection of helium in the spectra of two SNe Iax, SN~2004cs and SN~2007J. Note, however, that the classification of these two objects is questionable. While \citet{White15} have found these two events to be core-collapse (CC) Type IIb events rather than SNe Iax, \citet{Foley16} have contested these findings based on their own spectral and light-curve analysis.
Nevertheless, the scenario of He accreting progenitors has been strengthened by the detection of the luminous, blue progenitor system of the type Iax SN 2012Z \citep{McCully14a,Stritzinger15}, and is also consistent with the non-detection of the progenitor of SN~2014dt \citep{Foley15}. Detailed binary evolution calculations of \citet{Wang13} and \citet{Liu15} also support the possibility of such progenitor systems, although the latter authors note that this model is unlikely to be the most common progenitor scenario for SNe Iax. Moreover, \citet{Neunteufel17} suggested that explosions of He-accreting rotating magnetized CO WDs may be viable channels for faint and fast thermonuclear explosions like SNe Iax.

Beyond these possibilities, other progenitor models have been proposed in the literature, e.g. low-luminosity CC events emerging from 7-9 M$_{\odot}$, stripped-envelope stars, where part of the ejecta falls back to the central remnant \citep{Valenti09,Moriya10,Lyman13}; or a white dwarf merging with a neutron star or black hole \citep{Fernandez13,White15}. While these latter models seem to have much lower probability than the other ones, they cannot be completely ruled out at the moment.

We present in this paper a detailed spectral analysis of the Type Iax SN 2011ay, using the 1D radiative transfer code TARDIS \citep{Kerzendorf14}. We define a multi-layer atmosphere in the code, which allows us to carry out a method called ''abundance tomography'' \citep[see ][]{Stehle05}  in order to examine the physical properties and the chemical composition of the expanding ejecta in the early phases. Our main goal is to find a consistent spectral solution and check the validity of the different explosion scenarios.

In Section \ref{previous}, we give an overview of the target object of our study. In Section \ref{TARDIS}, we show how we use the TARDIS code to simulate the spectral evolution of SN~2011ay. In Section \ref{results}, we present the results of our spectral modelling, and compare the resulting physical picture of the ejecta with other models. Finally, we summarize our main findings in Section \ref{conc}.

\section{Target of study: SN~2011ay} \label{previous}
\subsection{Previous results}

SN~2011ay was discovered by \citet{Blanchard11} in NGC 2315 (Hubble-flow distance: d = 86.9 $\pm$ 6.9 Mpc) and classified as a 2002cx-like object by \citet{Silverman11}. Data of the SN were analyzed in the summarizing paper of \cite{Foley13} and, more thoroughly, by \cite{Szalai15}.
As has been shown in these papers, SN~2011ay belongs among the brightest ($M_\rmn{V, peak} \sim -18.4$ mag) and most energetic (i.e. most rapidly expanding) members of the Iax class.

Nevertheless, within the framework of the study of SN~2011ay, it has been revealed that the correct determination of the expansion velocities of SNe Iax is a challenging exercise \citep[as implied by the results of previous papers, see e.g.][]{Sahu08, McCully14b,Stritzinger14}.
\citet{Silverman11} and \citet{Foley13} derived the expansion velocity of SN~2011ay using the projected Doppler shift of the absorption minimum of the feature around $\sim$6200 \AA\, which was interpreted as due to Si\,\begin{small}II\end{small} $\lambda$6355, similar to Type Ia SNe. Assuming that the Si line forming region is at or very close to the photosphere, they considered this value, 5,700 km s$^{-1}$, as the photospheric velocity at maximum light in V band (labeled it $v_\rmn{peak}$). At the same time, \cite{Szalai15} showed that the absorption feature at $\sim$6200 \AA\ is very probably strongly affected by Fe\,\begin{small}II\end{small} $\lambda$6456 (and maybe also by some Co\,\begin{small}II\end{small}) even in the early phases. Since the velocity is usually estimated via fitting a single Gaussian profile to this region, such a 'quick-look' method can easily lead to  significant under- or overestimations of the true velocities of SNe Iax. Moreover, a direct connection between the position of the absorption minimum (usually called $v_\rmn{abs}$) and $v_\rmn{phot}$ is not generally possible in the cases where the lines are too weak or too strong \citep[e.g.][]{JB90,Dessart05a,Dessart05b,Blondin06}.

\cite{Szalai15} used SYN++ in order to identify lines and to determine the photospheric velocity of the ejecta of SN~2011ay.
The best-fit model they found gives a velocity of
$v_\rmn{phot}$ = 9,300 km s$^{-1}$ at V maximum, and consists of photospheric lines alone (their Model A). Nevertheless, an alternative model (Model B) also describes the spectra adequately; in this model, $v_\rmn{phot}$ at V maximum was set to 6,000 km s$^{-1}$, which is close to the value estimated by \citet{Silverman11} and \citet{Foley13}, but to get the proper fit it is necessary to allow the presence of detached line forming regions, which means that the minimum velocities of these regions, $v_\rmn{min}$, may be at higher velocities than $v_\rmn{phot}$.
Although Model A offers a definitely better solution, the non-uniqueness of the spectral model fitting shows that the velocities of SN~2011ay (and maybe those of most of SNe Iax) are rather ill-constrained.

\subsection{Observations and data} \label{spectroscopic}

The data sample of this study consists of ten optical spectra of SN 2011ay obtained between -3 and +19 days with respect to the moment of maximum brightness in B-band \citep[B-max; March 27, 2011,][]{Szalai15}.\footnote{Note that we use epochs relative to B maximum in the following, while \citet{Szalai15} used the time of V maximum as a reference.} Although even more late time spectra are available in the open acces databases, these are out of the reach of TARDIS, since the assumption of the photosphere is not fulfilled, as we discuss later in Sec. \ref{fitting} Six spectra were observed with the 9.2-m Hobby–Eberly Telescope (HET) Marcario Low Resolution Spectrograph \citep[spectral range: 4,100 - 9,500 \r{A},][]{Hill98}, published by \citet{Szalai15}, while four additional spectra were obtained with the 3-m Shane Telescope Kast spectrograph at the Lick Observatory \citep[spectral range: 3,500 - 10,000 \r{A},][]{Miller93}, published originally by \citet{Foley13} but analyzed in detail by \citet{Szalai15}. All the details of data reduction can be found in the two source papers. 

Before the analysis, all spectra have been corrected for redshift of the host galaxy using $z$ = 0.021 \citep[][]{Miller01} and the Milky Way interstellar reddening using $E(B-V)$ = 0.0694 mag. The latter value is based on dust emission maps from COBE/DIRBE and IRAS/ISSA \citep{Schlafly11} assuming $R_V$ = 3.1 and following the extinction law of \cite{Fitzpatrick07}.

\begin{figure}
\centering
\includegraphics[width=\columnwidth]{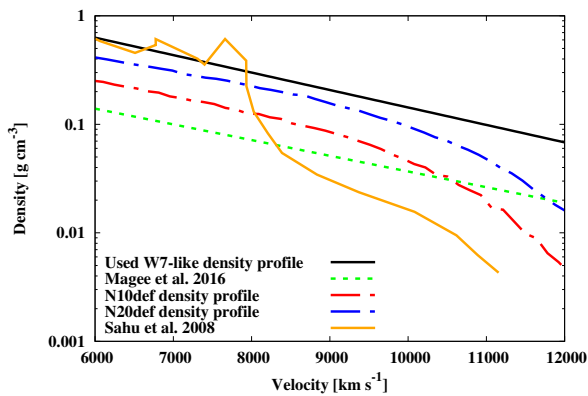}
\caption{\label{fig:density} Comparison of the exponential fit of the W7 density profile (black) used in this work to other density profiles in the literature, based on the analysis of Type Iax SN 2015H and SN 2005hk \citep[green and orange,][respectively]{Magee16,Sahu08} and pure deflagration models \citep[red and blue,][]{Fink14}.}
\end{figure}

\begin{figure*}
\centering
\includegraphics[width=130mm]{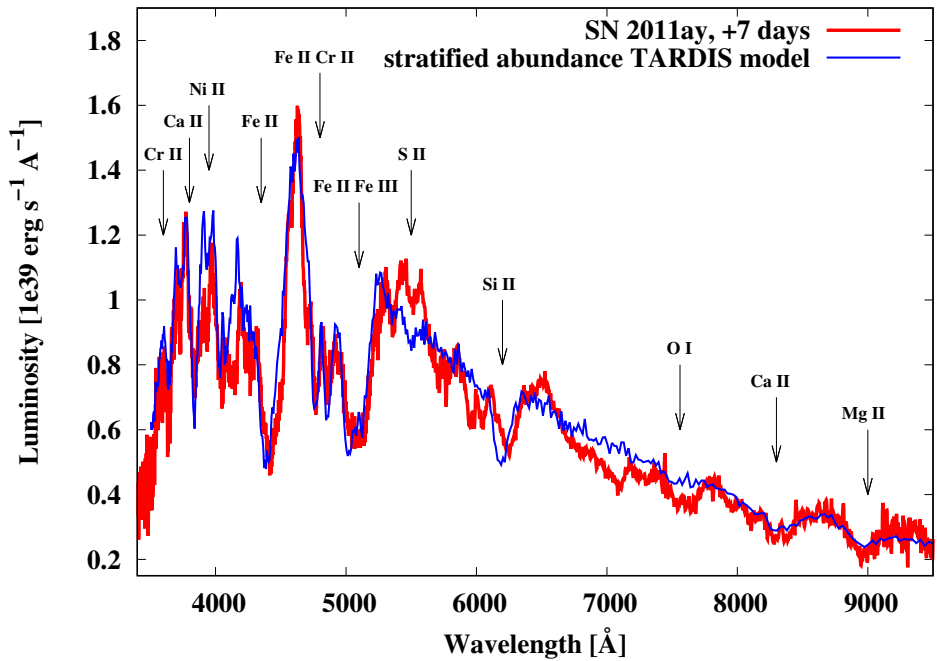}
\caption{\label{fig:ions} The spectrum of SN 2011ay at the epoch +7 days with respect to B-maximum (red) and the synthetic spectrum from the fit of the velocity-dependent  model (blue). The black arrows show the position of prominent absorption features and the related ions.}
\end{figure*}

\section{METHODS} \label{TARDIS}


The Monte Carlo (MC) radiative transfer code TARDIS \citep{Kerzendorf14} assumes the supernova consists of an opaque core emitting a black-body continuum through a mostly transparent homologously expanding ejecta. The spectrum emerging from the ejecta is calculated by sending indivisible photon packets, all representing a bundle of identical photons, through the expanding ejecta and following their interaction with matter. The frequencies of these MC packets are randomly selected from an initial blackbody distribution at the opaque inner boundary, the temperature of which is calculated from the given emergent luminosity according to the Stefan-Boltzmann law.
 
In TARDIS, the model atmosphere is constrained by an inner and an outer boundary (both defined by their velocities in the homologously expanding atmosphere) and divided into multiple cells (spherical shells). For each of these, the density and elemental abundances must be specified. The model ejecta can be defined by the user in a custom velocity grid. During their propagation through this grid, the packets may interact with matter via Thomson scattering (by electrons) or bound-bound transitions (by atoms/ions, calculated via the Sobolev-approximation). These interactions might change the frequencies and directions of photon packets, but the energy in the comoving frame of a packet is conserved during all the interactions. 

The whole TARDIS calculation consists of an iterative sequence of radiative transfer simulations. At the end of a propagation, the escaping packets are used to calculate the synthetic spectrum and to prepare the next iteration cycle. Based on the flight history of the photon packets, the ionization and excitation conditions belonging to each of the grid cells are derived from the MC radiation field assuming local thermodynamic equilibrium. The blackbody temperature at the inner boundary is also recalculated, matching to the value of the emergent luminosity. The new temperature-depending conditions of the model ejecta are adopted for the next computational cycle, thus the iteration improves the temperature-dependent conditions to a more consistent model.

On the whole, the input parameters are the emergent luminosity ($L_{\rmn{e}}$), the time since explosion ($t_{\rmn{exp}}$), the density and chemical composition of each cell, and the velocity boundaries of the model. 

In spite of the relatively complex approach, the code allows rapid modeling of spectra: the usual execution time is 1-2 minutes on a typical desktop or laptop computer. These attributes make TARDIS an excellent option to fit the observed spectra of thermonuclear SNe and discover their physical and chemical properties.

\subsection{Fitting methods} \label{fitting}

As the SN expands, the ejecta becomes thinner and thinner and the photosphere recedes to deeper and deeper layers. This means that spectra from the early and late epochs probe different regions of the ejecta, thus allowing the construction of a velocity-dependent abundance profile from a spectral time series of an observed supernova \citep[abundance tomography,][]{Stehle05}.


Following the strategy described below, we fit ten photospheric spectra of SN~2011ay (listed in Table \ref{tab:log}) with \begin{small}TARDIS\end{small}, getting insight into the composition above the photosphere. The last analyzed spectrum obtained at $\sim$32 days after the explosion, which approximately corresponds to the time limitation of the TARDIS model conditions \citep[][]{Kerzendorf14,Magee16}.
TARDIS requires information about the structure of the SN ejecta. The density profiles of the atmospheres of SNe Iax, originating from different model calculations in the literature, vary on a wide scale; however, an exponential decrease of the density toward the outer regions is a common assumption:
\begin{equation}
\rho (v,t_{\rmn{exp}}) = \rho_0 \cdot \left(\frac{t_{\rmn{exp}}}{t_{\rmn{ref}}}\right)^{-3} \cdot \exp\left({-\frac{v}{v_0}}\right),
\end{equation}

\noindent where $t_{\rmn{ref}}$ is the reference time, $\rho_{\rmn{0}}$ is the central density at $t_{\rmn{ref}}$, $v_0$ is the e-folding velocity and $t_{\rmn{exp}}$ is the time since explosion. As a first step, we set the values of the parameters ($\rho_{\rmn{0}} = 9 \cdot 10^{-9}$ g cm$^{-3}$, $t_{\rmn{ref}}$ = 1 day, $v_0$ = 2670 km s$^{-1}$) to be consistent with the exponential fit of the well-known W7 explosion model \citep{Nomoto84} sometimes applied in the cases of "normal" SNe Ia.
We used this density profile during the whole modeling process in order to reduce the number of free parameters. We note, however, that the W7 explosion model does not represent either SNe Iax or deflagrations with bound remnants in general; thus, uncertainties could arise from this assumption. Nevertheless, looking through the literature (see Fig. \ref{fig:density}), either higher or lower values of $\rho_{\rmn{0}}$ have been used to describe the atmospheres of SNe Iax, see e.g. \citet{Sahu08} and \citet{Magee16}. Since SN 2011ay is one of the most luminous SNe Iax ($M_{\rmn{V}}$ $\sim$ -18.4 mag) and its expansion velocity is close to the values of normal Type Ia SNe, assuming a W7-like density profile seems to be a reasonable approximation.

The total mass of the ejecta according to the adopted density profile and the appropriate velocity range is $\sim$ 0.95 M$_{\odot}$. This is higher than the M$_{ej}$ = 0.8 M$_{\odot}$ in \cite{Szalai15}, but their method enables only a rough estimation, while our value comes from a fixed density profile. Thus, the comparison of the ejecta mass values does not indicate any inconsistency.

By definition, the inner boundary of the TARDIS model atmosphere is the bottom of the computation volume, where a blackbody radiation field is assumed. Hereafter, we will also refer to this boundary as the photospheric velocity. 
The outer boundary of the TARDIS computational domain has to be sufficiently far from the inner boundary, so that the thin, outer part of the atmosphere does not affect the spectrum. Thus, the outer boundary is fixed at 5000 km s$^{-1}$ above the photosphere, where the density of the atmosphere is lower by orders of magnitude depending on the density profile.

The values of the emergent luminosity at the observed epochs, $L_{\rmn{e}}$, as well as the date of explosion, $t_{\rmn{0}}$ = 2\,455\,633.0 $\pm 1.5$ JD, are adopted from \citet{Szalai15}. The uncertainties of epochs and luminosities result in significant variations in the output spectra. To avoid inconsistencies during the fitting, we fix the value of $t_{\rmn{exp}}$ at each epoch. At the same time, we allow the luminosities to vary within an average uncertainty, $5 \times 10^{41}$ erg s$^{-1}$, estimated from the uncertainties of $L_{\rmn{e}}$ given by \citet{Szalai15}.

\begin{figure}
\centering
\includegraphics[width=\columnwidth]{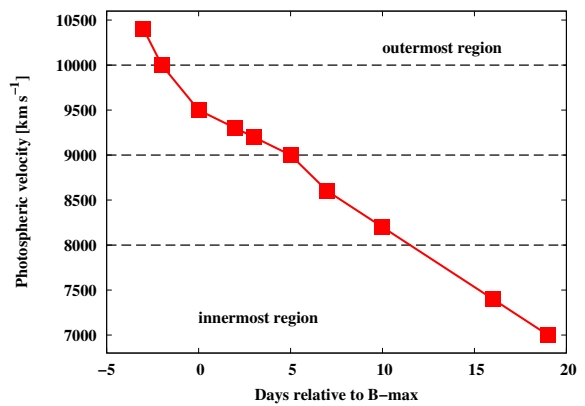}
\caption{\label{fig:vphot} The photospheric velocities of SN 2011ay originating from our calculations with TARDIS. The dashed lines show the velocity grid applied in our models.}
\end{figure}

We manually fit TARDIS models to ten spectra of SN~2011ay obtained between -3 and +19 days relative to B-maximum. We use a velocity grid with steps of 1000 km s$^{-1}$ (see Fig. \ref{fig:vphot}), where the mass fraction of the elements are fitting parameters in each cell. 
A main aspect during our work is to ensure a self-consistent and realistic approach to an expanding SN ejecta as much as possible. 
Since the mass fractions of the elements do not change in time (aside from the nuclear decay of radioactive elements), we make an effort to find one set of abundances that can be adequately used for modeling all epochs. More precisely, since we generally use the stratified modeling method, it means that the mass fraction of a given element can be different from cell to cell, but the composition of a given cell (thus, the total abundance of a given element, too) will be the same at every epoch in our final solution. This is not true for iron, cobalt and nickel mass fractions, which are altered according to the $^{56}$Ni decay. For this purpose, we assumed that the whole amount of cobalt and nickel consist of $^{56}$Co and $^{56}$Ni isotopes. Although, these two isotopes cover the majority of cobalt and nickel in the SN ejecta, non-negligible fractions of stable IGE are also produced, thus, our assumption leads to a further simplification.

For determining the mass fractions of elements in a certain layer, we use the spectrum obtained at that epoch when $v_\rmn{phot}$ is close to the velocity of this layer (see again Fig. \ref{fig:vphot}). 
This means that the mass fractions above 10,000 km s$^{-1}$ come from fitting the two earliest spectra, obtained at -3 and -2 days. In the next step, the mass fractions in the layer between 9,000 and 10,000 km s$^{-1}$ are determined by fitting the spectra taken at 0, +2, and +5 days, while, finally, the mass fractions between 8,000 and 9,000 km s$^{-1}$ are specified based on the spectra obtained at +7 and +10 days. If we find any major discrepancy between the observed and synthetic spectra at any points, we step back and try to find another solution at the earlier epoch. Although this method does not necessarily give the best fit at every epoch, it guarantees the self-consistency of our model.

\subsection{Used chemical elements in TARDIS model} \label{elements}

Building up the TARDIS model atmosphere, we only use elements forming significant spectral features at the observed epochs: O, Mg, Si, S, Ca, Cr, Fe, Co, and Ni (see Fig. \ref{fig:ions}). Note that other elements (e.g. Ne, Na), which do not show strong spectral features, may be also present in the atmosphere of an SN Iax. These ''hiding'' elements might have minor effect on the observed spectrum, but, even if their mass is significant in the ejecta, their amounts would be highly uncertain in our analysis. 

\begin{figure}
\centering
\includegraphics[width=\columnwidth]{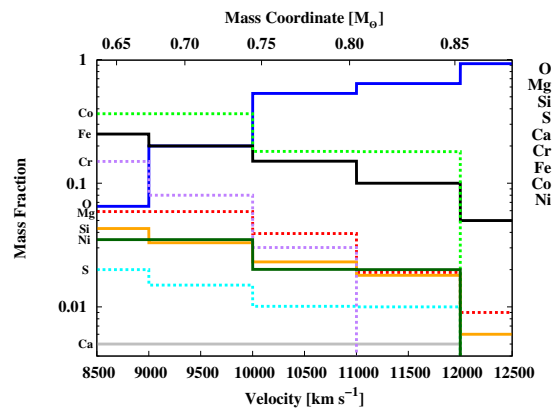}
\caption{\label{fig:VD} The stratified abundance profile from the TARDIS fitting process of all the SN 2011ay spectra. The dashed lines show the photospheric velocities at each epoch. The Fe/Co/Ni mass fractions are from the epoch +7 days with respect to B-maximum. The mass fraction of the oxygen goes up to 0.65 between 11,000 and 12,000 km s$^{-1}$ and reaches 0.93 above 12,000 km s$^{-1}$.}
\end{figure}

We have found that there is a considerable mass fraction (0.10 - 0.50, depending on the epoch and the layer) above 10,000 km s$^{-1}$, which cannot be covered by the elements used in our TARDIS model. We find that involving any other elements decreases the goodness of the fit. The presence of this excessive mass fraction may be the result of using an inappropriate density profile (see Fig. \ref{fig:density}). Another possibility is that elements with extremely optically thick lines (e.g. oxygen), whose amount is only roughly estimated, still cover these fractions, since these lines are close to being saturated. Additionally, it could be an effect caused by the above mentioned ''hiding'' elements (Ne, Na etc.), at least partially. As we have also tested, the atoms of these elements, despite their increased mass fraction, do not interact with the photon packets in our TARDIS model. Therefore, the estimation of the amount of one or more ''hiding'' elements is completely uncertain. Thus, as a solution -- and a further simplification --, we choose to ''fill up'' the excessive mass fraction only with oxygen in such cases. This method leads to overestimate \textit{only} the fraction of oxygen, but prevents including unreal components in the model. Based on our detailed tests, using an artificially increased amount of oxygen does not change the model spectra, since the only O \begin{small}I\end{small} line at $\sim$7500 \r{A} becomes optically thick at even a relatively moderate amount of oxygen, while the physical parameters of the ejecta do not seem to allow the formation of higher-excitation lines of O \begin{small}I\end{small}.

Since carbon is one of the key element in thermonuclear supernovae, we investigated the possible use of it in the TARDIS model atmosphere in depth.

Carbon is not formed in the SN atmosphere as a nuclear burning product, but comes from the original matter of the progenitor WD. Thus, it has been argued that the detection of carbon can be used to constrain the progenitor and explosion mechanism of a supernova. C\,\begin{small}II\end{small} $\lambda$6580 was detected in several SNe Iax \citep{Chornock06,Foley10a,McClelland10,Parrent11,Thomas11b}; moreover, \citet{Foley13} claimed that every SN Iax may show carbon features in their spectra before or around maximum light. The pure deflagration explosion scenario also predicts significant amount of carbon, even in the inner region of the atmosphere.

\begin{figure}
\centering
\includegraphics[width=\columnwidth]{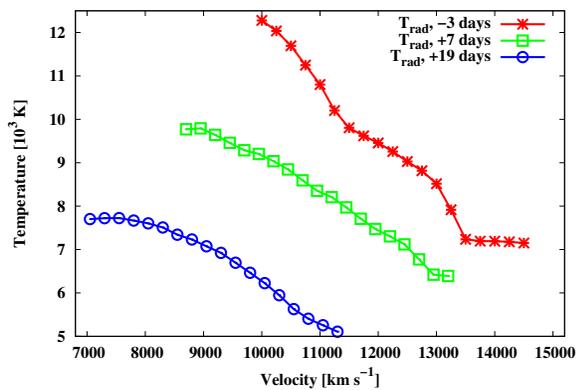}
\caption{\label{fig:t_rad} Radiation temperature in the velocity cells of the stratified TARDIS model at the epochs -3, +7 and +19 days with respect to B-band maximum}
\end{figure}

\begin{figure}
\centering
\includegraphics[width=\columnwidth]{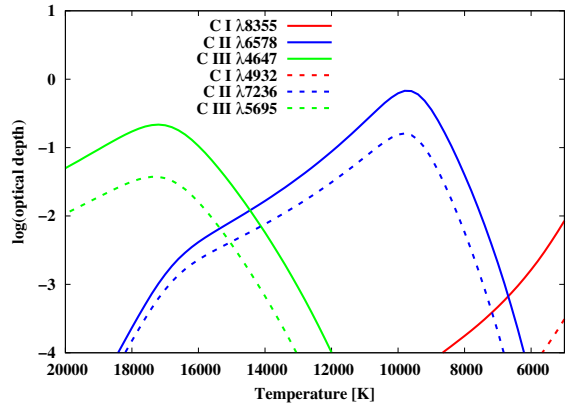}
\caption{\label{fig:opt_depth} Estimated optical depths for strong carbon lines as function of temperature at B-max, assuming a carbon mass fraction of 0.10 above 10,000 km s$^{-1}$. The calculation is based on the Eq. 2. of \citet{Hatano99} assuming LTE conditions.}
\end{figure}

As a first step, we calculated the expected optical depth values of the strong carbon lines based on the method of \cite{Hatano99}. From our TARDIS simulations, we can extract the radiation temperature \citep[for further description see][]{Mazzali93} in the SN ejecta. We find that the temperature changes between 5,000 and 12,000 K in the ejecta during the observed time frame (see Fig. \ref{fig:t_rad}). According to these simple LTE calculations, only singly ionized carbon lines having optical depths $\tau > 0.01$ are expected to appear in the observed spectra of SN~2011ay (see Fig. \ref{fig:opt_depth}).

\begin{figure}
\centering
\includegraphics[width=\columnwidth]{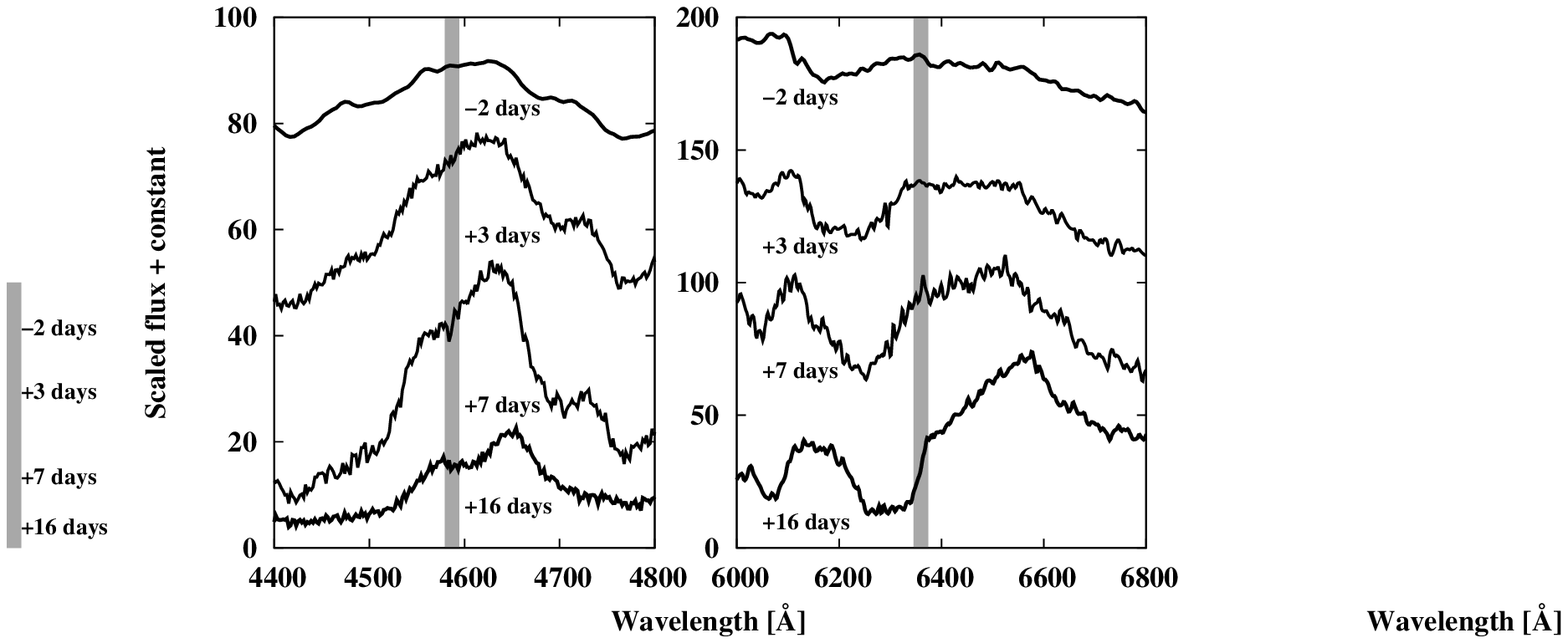}
\caption{\label{fig:carbonara} The observed spectra of SN 2011ay obtained at the labeled time. The gray lines show the position of singly-ionized carbon lines $\lambda$4746 and $\lambda$6578 respectively, redfshifted with 10,000 km s$^{-1}$.}
\end{figure}

The most prominent lines of C\,\begin{small}II\end{small} expected to appear at $\lambda$4746 (this feature is caused by several weaker lines, thus, it does not appear in Fig. \ref{fig:opt_depth}), $\lambda$6578 and $\lambda$7234.
No strong feature can be seen in any of the observed spectra at these wavelengths (see Fig. \ref{fig:carbonara}). It is true, however, that some of the weaker spectral features close to the expected wavelengths of the C\,\begin{small}II\end{small} lines may be due to carbon, especially if the pseudo-continuum runs above the
model continuum of our TARDIS models (see Sec. \ref{syn++_tardis}). Because of the
limitations of our TARDIS modeling, we cannot get reliable constraints on the
amount and distribution of carbon in the atmosphere of SN~2011ay. Therefore,
we have not included carbon in our TARDIS models. As a result, we refrain from setting constraints on the amount and
distribution of carbon and in the absence of strong absorption features we also eliminated it from our TARDIS model fitting process.

\begin{table*}
	\centering
	\caption{Log of the analyzed spectra of SN~2011ay, the time with respect to B-maximum \citep[$T_\rmn{max}$ = 2,4556,46.6 from][]{Szalai15} and the associated TARDIS parameters: the time since explosion, the emergent luminosity and the velocity of the inner boundary. For further description see Sec. \ref{fitting}. }
    \label{tab:log}
    \begin{tabular}{cccccc}
		\hline
		Observatory&\begin{tabular}{@{}c@{}}JD \\-2\,450\,000\end{tabular}& \begin{tabular}{@{}c@{}}$t_\rmn{epoch}$ \\ (days)\end{tabular} & \begin{tabular}{@{}c@{}}$t_\rmn{exp}$ \\ (days)\end{tabular}&log ($L_\rmn{e}$)&\begin{tabular}{@{}c@{}}$v_\rmn{inner}$ \\ (km s$^{-1}$)\end{tabular}\\
		\hline
		HET & 5643.2 & -3.4 & 10.2 & 42.54 & 10,400\\
		HET & 4644.2 & -2.4 & 11.2 & 42.58 & 10,000\\
		HET & 5646.2 & -0.4 & 13.2 & 42.64 & 9,500\\
        HET & 5648.2 & +1.6 & 15.2 & 42.66 & 9,300\\
        Lick & 5649.2 & +2.6 & 16.2 & 42.69 & 9,200\\
        HET & 5651.2 & +4.6 & 18.2 & 42.70 & 9,000\\
        Lick & 5653.2 & +6.6 & 20.2 & 42.71 & 8,600\\
        Lick & 5656.2 & +9.6 & 23.2 & 42.61 & 8,200\\
        Lick & 5662.2 & +15.6 & 29.2 & 42.47 & 7,400\\
        HET & 5665.2 & +18.6 & 32.2 & 42.43 & 6,800\\
		\hline
	\end{tabular}
\end{table*}

\begin{table}
	\caption{The treatments of physical processes in our TARDIS model, and the status of the physical parameters in our fitting process. For further description see Sec. \ref{fitting} and \citet{Kerzendorf14}.}
    \label{tab:parameters}
	\begin{tabular}{l}
        ionization: lte \\
    	excitation: lte \\
    	radiative rates: dilute-blackbody \\
    	line interaction: macroatom \\
	\end{tabular}
	\begin{tabular}{lcc}
        \hline
        General properties & & \\
        $t_{\rmn{exp}}$ & fixed & \citet{Szalai15} \\ 
        $L_{\rmn{e}}$ & fitting & --\\
        \hline
        Computational domain & & \\
        $v_\rmn{inner}$ & fitting & --\\
        $v_\rmn{outer}$ & fixed & 5,000 km s$^{-1}$ above $v_\rmn{inner}$\\
        \hline
        Density profile & & \\
        $\rho_{\rmn{0}}$ & fixed & exponential fit of W7 profile\\
        $v_0$ & fixed & exponential fit of W7 profile\\
        $t_{\rmn{ref}}$ & fixed & exponential fit of W7 profile\\
        \hline
    \end{tabular}
\end{table}

\begin{figure*}
\centering
\includegraphics[width=16cm]{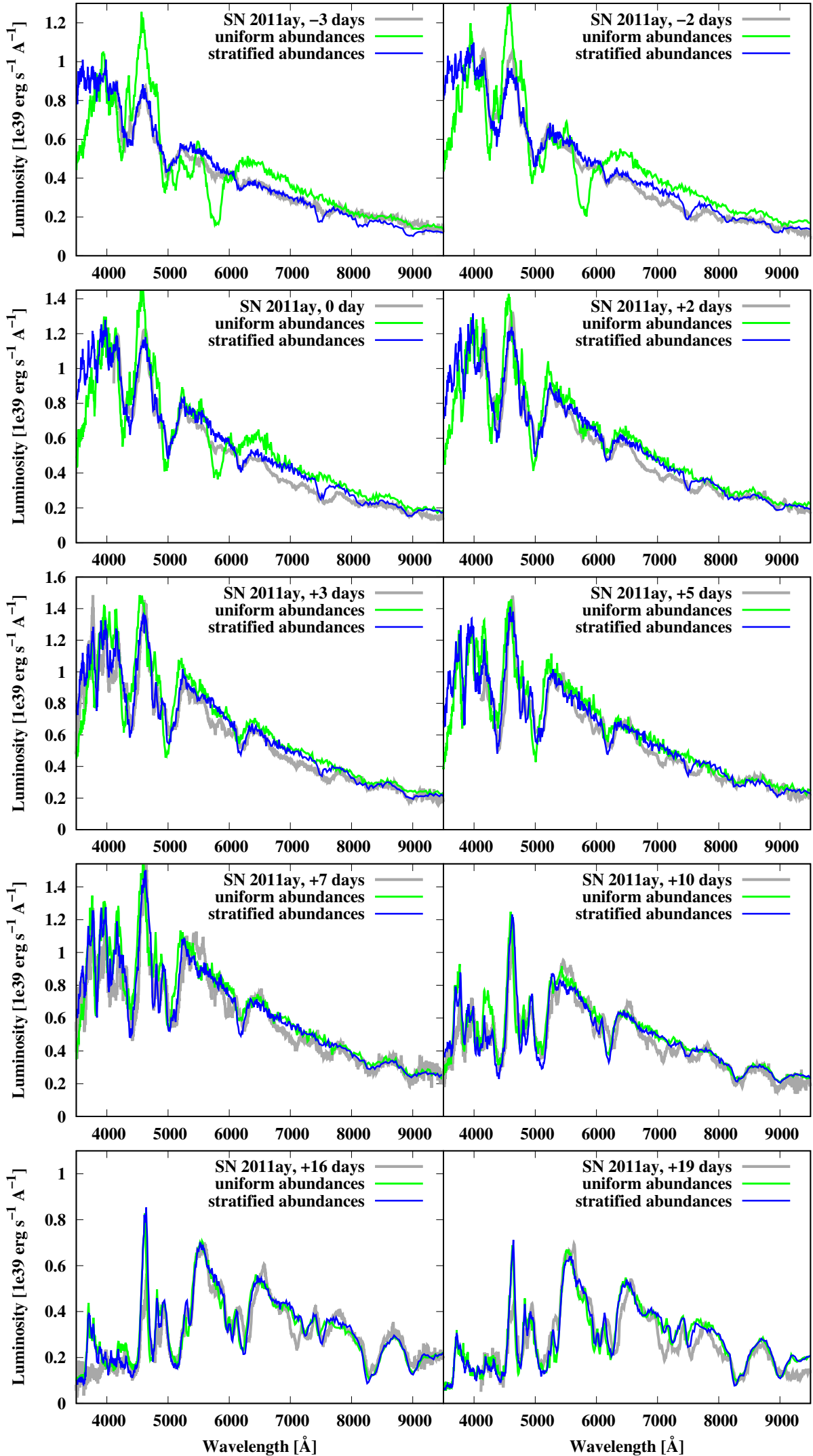}
\caption{\label{fig:spectra} The observed spectra (grey) of SN 2011ay obtained between -3 and +19 days with respect to B-max, compared to the best-fit TARDIS models assuming a stratified (blue) and a uniform (green) abundance profile.}
\end{figure*}

\section{Results and discussion} \label{results}

\subsection{TARDIS fitting of SN 2011ay} \label{tardisfitting}

The resulting velocity-dependent, stratified abundance-structure of the TARDIS model is illustrated in Fig. \ref{fig:VD}. Note that the Ni/Co/Fe abundances change in time according to the $^{56}$Ni decay sequence, while the mass fractions of the other elements are fixed. In the best-fit \begin{small}TARDIS\end{small} model, the highest mass fraction belongs to oxygen; however, the amount of oxygen is probably overestimated in our model, because of the applied fitting method (see Sec. \ref{elements}). At 7 days after B maximum, the mass fraction of cobalt, which is the second most abundant element, reaches 0.35 below 9,000 km s$^{-1}$, but it decreases toward the outer ejecta. Iron shows a similar trend with mass fractions of 0.15-0.25 in the inner parts and a continuous decrease toward the outer regions.

Intermediate mass elements appear with smaller amounts than IGEs; all of their mass fractions are below 0.06. Mass fractions of magnesium and silicon also show a decreasing trend (from 0.06 to 0.01 and from 0.04 to 0.005, respectively) toward the higher velocities, while calcium mass fraction is constantly 0.005 in the analyzed velocity space. Sulphur has only a moderate (0.005-0.02) mass fraction in the whole velocity space of the model and it changes together with the mass fraction of silicon.

\begin{figure*}
\centering
\includegraphics[width=130mm]{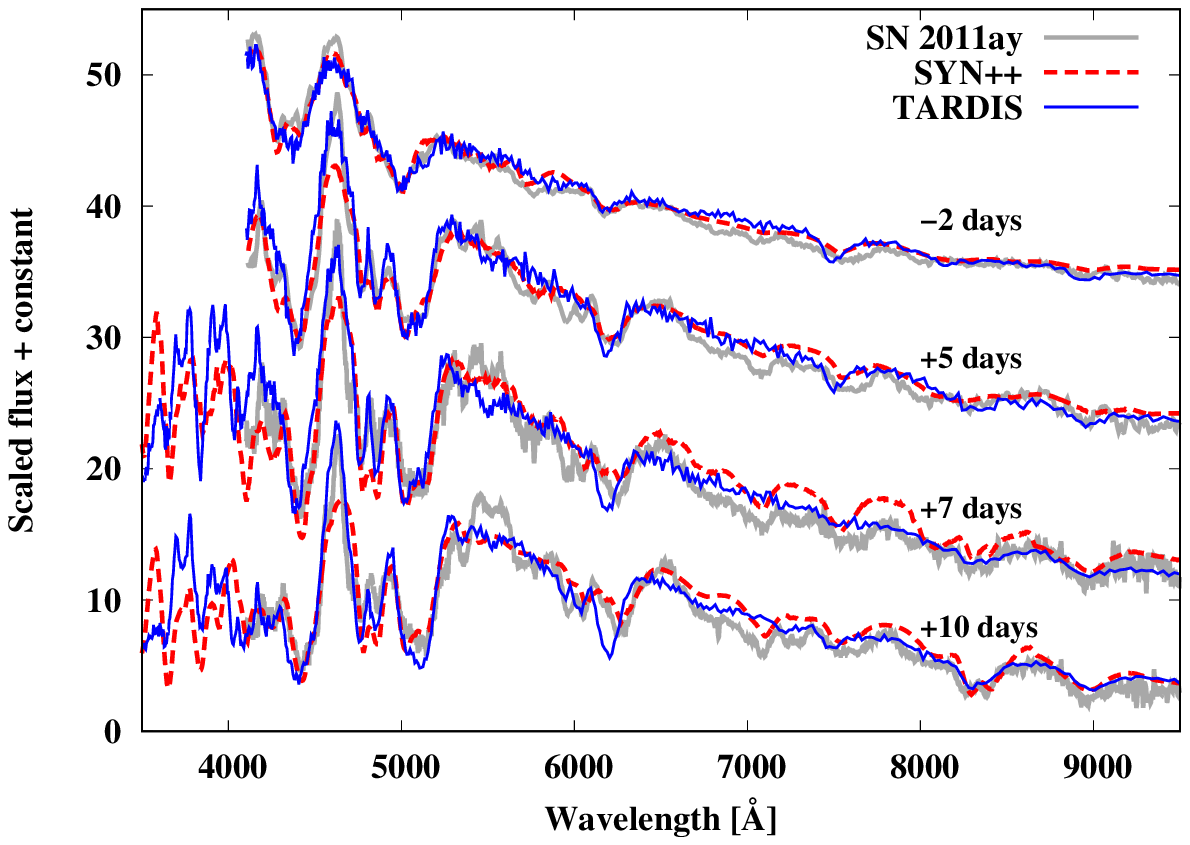}
\caption{\label{fig:syn++} The observed spectra of SN~2011ay (grey), compared to the synthetic spectra of the SYN++ model (red) of \citet{Szalai15} and our best-fit TARDIS model (blue).}
\end{figure*}

This abundance distribution is clearly able to fit both the global shape of the spectra and most of the spectral features at all the observed epochs (see Fig. \ref{fig:spectra}). As we mentioned in Sec. \ref{elements}, all of the adopted ions appear in the output spectra with characteristic features. There are a few systematic discrepancies, which are discussed in the next subsection.

Assuming a stratified abundance-structure of SN~2011ay may be questionable since pure deflagration models, which seem to serve as the best explanation for several Iax explosions (see Section \ref{introduction}), predict a well-mixed ejecta. Thus, we need to investigate whether a single, constant abundance-structure is able to reproduce the whole spectral series of SN~2011ay or not.
In order to do this, we also create synthetic spectra with uniform abundances to test the fittings compared to the stratified abundance model. For the uniform abundance model, we fit the spectrum obtained at +7 days with constant mass fractions and adopt this distribution for the other epochs considering the radioactive decay of  $^{56}$Co and $^{56}$Ni isotopes.

The resulting synthetic spectra of the two different TARDIS models (uniform abundances and stratified abundances) are plotted in Fig. \ref{fig:spectra}. Although the stratified abundance structure is much more complex, the uniform abundance model also fits the spectra obtained after B-maximum well. At the same time, we are not able to achieve such good fits in the cases of pre- and near-maximum spectra, especially because of the extreme strength of Fe\,\begin{small}III\end{small} absorption features at the early epochs. The strength of these features could be reduced in the model by setting lower emergent luminosity or modifying $t_\rmn{exp}$. However, in this case, our model would be in conflict with the results of the photometric measurements. Note that varying the abundances in the uniform model from epoch to epoch could lead to an equally good fit compared to the stratified model, but this approach would destroy one of our most important intentions, the self-consistency of our modeling process.

Note again that we use a fixed density profile and this leads to a reasonable fit with stratified abundances but not with constant abundances. It should also be noted that this density profile is quite different from the previously published density profiles of deflagration models assuming bound remnants \citep[see e.g.][]{Kromer13,Fink14}. It cannot be excluded that it is this difference that drives our TARDIS fits to strongly favour a stratified abundance profile.

\subsection{ Comparison of SYN++ and TARDIS models} \label{syn++_tardis}

In the following, we briefly compare our best-fit \begin{small}TARDIS\end{small} model to the SYN++ fits of \cite{Szalai15}. Since the inner boundary of our \begin{small}TARDIS\end{small} models are closer to the $v_\rmn{phot}$ values of their Model A (see Sec. \ref{previous}) at the same epochs, we choose this \begin{small}SYN++\end{small} model set for comparison.

All the main features of the spectra of SN 2011ay can be fit using either \begin{small}SYN++\end{small} or \begin{small}TARDIS\end{small} (see in Fig. \ref{fig:syn++}). Nevertheless, both codes have difficulties to fit some features, like the depression of the flux beyond 7,500 \r{A} at -3 days. This phenomenon cannot be explained by the line forming of any specific elements. It follows that these failings may come from the common limitation of the synthesis codes, like the assumption of an underlying blackbody continuum. It is apparent, especially on the +7 days spectrum, that the pseudo-continuum of the TARDIS model spectrum runs below that of the SYN++ spectrum redward of 6500 \r{A}. This makes the identification and modeling of the weaker features uncertain in this wavelength regime (see also Sec. \ref{elements} regarding the presence/absence of carbon features).

The elements used in our \begin{small}TARDIS\end{small} model atmosphere generally agree with those used in the \begin{small}SYN++\end{small} model of \citet{Szalai15}, except the lack of sodium, which causes no significant absorption features in our \begin{small}TARDIS\end{small} models. Na\,\begin{small}I\end{small} requires very low temperature (< 6,000 K) to form its strong lines in the optical spectral range, but it is outside the temperature range of our \begin{small}TARDIS\end{small} models (see Fig. \ref{fig:t_rad}). Regarding the line forming ions, the only significant difference is the presence of the extended O\,\begin{small}II\end{small} line around $\sim$7,000 \r{A}, which occurs in the \begin{small}SYN++\end{small}, but not in \begin{small}TARDIS\end{small} spectra.

Taking a look at the synthetic spectra created from the \begin{small}TARDIS\end{small} model, the spectral feature of Si\,\begin{small}II\end{small} $\lambda$ 6355 \r{A} (mentioned in Sec. \ref{previous}) is properly fit up to +5 days. At later epochs, the synthetic spectral feature becomes blueshifted in our TARDIS model compared to the observed fluxes. This means that the amount of Si\,\begin{small}II\end{small} is overestimated in the outer atmosphere, either due to differences in the ionization state or an excessive mass fraction of silicon. The latter explanations seem to be in conflict with the findings from the earlier epochs, when the higher silicon mass fractions are necessary to fit the $\lambda$6355 \r{A} feature. Similar trends are observed in the case of the strong iron features. A possible origin of this contradiction may be the application of the W7 density profile without any specialization referring to SNe Iax.

\subsection{Comparison to pure deflagration models}

As mentioned in Sec. \ref{introduction}, pure deflagration models are promising for explaining the observational attributes of SNe Iax. Both the calculated $^{56}$Ni mass and the total ejecta mass \citep[$M\rmn{_{Ni}}$ = 0.22 $M\rmn{_{\odot}}$ and $M\rmn{_{ej}}$ = 0.8 $M\rmn{_{\odot}}$,][]{Szalai15} suggest that SN~2011ay may have exploded in a pure deflagration of a C/O WD.
Therefore, we compare the observed spectra of SN 2011ay with the results of the 3D hydrodynamic simulations published by \cite{Fink14}. In this work, the initial deflagration strength is parameterized by a varying number of ignition spots that are placed near the center of a M$\rmn{_{Ch}}$ WD. The nomenclature of the model grid, which is adopted also in our study, corresponds to the number of ignition spots (i.e. N20def means that there are 20 spots in the model). 
The hydrodynamic simulations, on which the calculations of the nucleosynthesis yields and the synthetic observables are based, follow the explosion up to $t\rmn{_{exp}}$ = 100 s. \citet{Fink14} showed that the number of ignition spots strongly correlates with the released energy and the luminosity of the SN. In the created model grid, the models with ignition spots up to $\sim$100 leave a bound remnant behind with a mass of $M\rmn{_{b}}$ > 0.1 $M\rmn{_{\odot}}$. The authors show that some of their deflagration models with bound remnant fit the light curves and spectra of some Type Iax SNe reasonably well.

Comparing the maximum brightness values in B,V, and R bands (-18.15$\pm$0.17, -18.39$\pm$0.18, -18.60$\pm$0.17 mag, respectively) to those originating from the synthetic model grid, we find that N10def (-17.95, -18.38, -18.36) and N20def (-18.24, -18.68, -18.73) models are comparable to the observed values of SN 2011ay.
These models assume 10 and 20 ignition spots, respectively, which trigger the explosion of a M$\rmn{_{Ch}}$ WD with an initial composition of X(C) = 0.475, X(O) = 0.50 and X(Ne) = 0.025. The central density is set to $\rho\rmn{_{c}}$= 2.9 $\times$ 10$^{9}$ g cm$^{-3}$, while the temperature profile is assumed as constant at T = 5 $\times$ 10$^{5}$ K.

\begin{figure}
\centering
\includegraphics[width=\columnwidth]{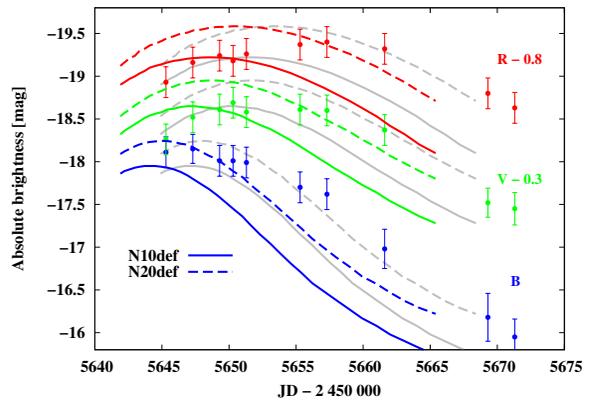}
\caption{\label{fig:photometry} Standard BVR absolute magnitudes of SN 2011ay compared to the synthetic light curves calculated from the N10def and N20def pure deflagration models published by \citet{Fink14}. The colored light curves belong to the explosion date obtained from \citep{Szalai15}, while the grey light curves are shifted by +3 days for better fitting.}
\end{figure}

In the next step, we compare the shapes of the observed and synthesized BVR light curves. At first, the explosion date is chosen as $t_0$ = 2\,455\,633.0($\pm 1.5$) JD adopted from \citet{Szalai15}; using this value, the evolution of our \begin{small}TARDIS\end{small} model spectra seem to be in good agreement with the observations. 
As can be seen in Fig. \ref{fig:photometry}, neither N10def nor N20def models are able to consistently fit the observed magnitudes; at least, using the adopted value of $t_0$, we get conspicious differences. Assuming that the estimated uncertainty of $t_0$ (1.5 days) published by \citet{Szalai15} may be too optimistic, we are allowed to modify the value of $t_0$ by three days (choosing 2\,455\,636.0 JD instead of 2\,455\,633.0 JD), which gives a better agreement. Nevertheless, these fittings are also controversial. While B magnitudes seem to be consistent with the N20def model, V and R magnitudes follow the N10def rather than the N20def model. Moreover, the synthetic light curves evolve too fast in the latter two bands compared to the observations.
As we have already noted in Sec. \ref{introduction}, this seems to be a general problem of pure deflagration models of SNe Iax, which could be caused by an excessive ejecta-mass at constant $^{56}$Ni mass increasing the opacity of the ejecta.

\begin{figure*}
\centering
\includegraphics[width=11cm]{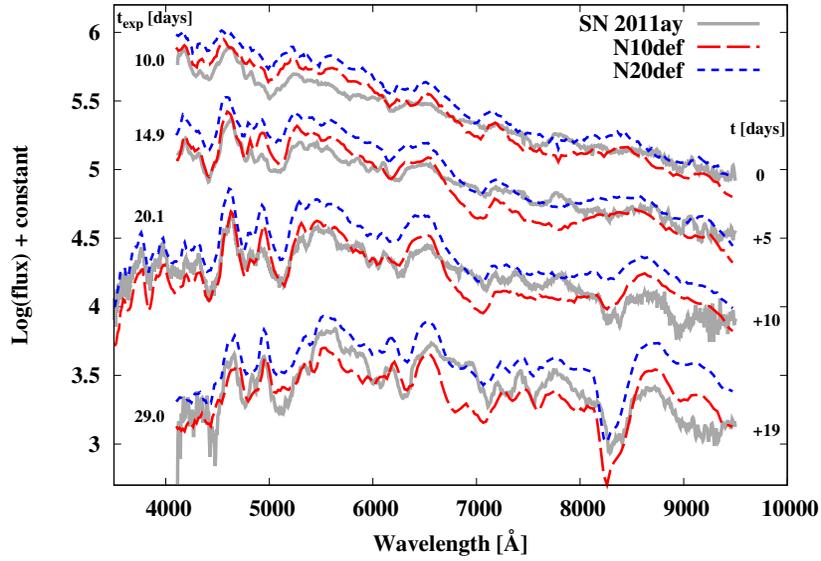}
\caption{\label{fig:fink_spectra} Comparison between the observed spectra of SN 2011ay (grey) and the synthetic spectra originating from N10def (red) and N20def (blue) models \citep{Fink14}. The meaning of the two time axes are the days since the explosion assumed by the deflagration models (left) and the days with respect to B-max in the cases of the observed spectra (right). }
\end{figure*}

We also compare the synthetic N10def and N20def model spectra to the observed spectra of SN 2011ay (Fig. \ref{fig:fink_spectra}). According to the photometric comparison, we use here the later date of explosion ($t_0$ = 2\,455\,636.0 JD) to choose the model spectra. Before and around the time of B-band maximum, both of the models show poor agreement with the observed data; however, the main observed spectral features, like the strong iron absorption at $\sim$5,000 \r{A} or the P Cygni profile of Si\,\begin{small}II\end{small} $\lambda$6355, can be also recognized in the synthetic spectra. While the N20def model is apparently too luminous in this phase, the fainter N10def model is closer to the observed flux level considering the whole spectral range. Note that a model with fewer ignition spots (e.g. N5def), which results in lower flux levels, might also be able to fit the spectra obtained at the early epochs.

After B-max, the fitting results using the N10def model become better, especially between +5 and +10 days, when the synthetic spectra show a remarkably decent match below $\sim$6,500 \r{A}; both the strong features formed by iron ions and the narrow features of other elements are fit successfully. At the same time, the flux level between 6,500 and 8,000 \r{A} is still underestimated by the N10def model. The N20def model is able to reproduce the observed flux level in this region, but only some of the observed absorption features (mostly oxygen lines) appear in the synthetic spectrum. A further discrepancy is the conspicuous lack of the strong absorption of Mg\,\begin{small}II\end{small} at $\sim$8,800 \r{A} in the synthetic spectra before and around maximum light.
Choosing different parts of the spectrum at +19 days, sometimes the N10def, sometimes the N20def model shows a better fit, which raises the possibility that a transition model would be the best solution. 

\begin{figure*}
\centering
\includegraphics[width=11cm]{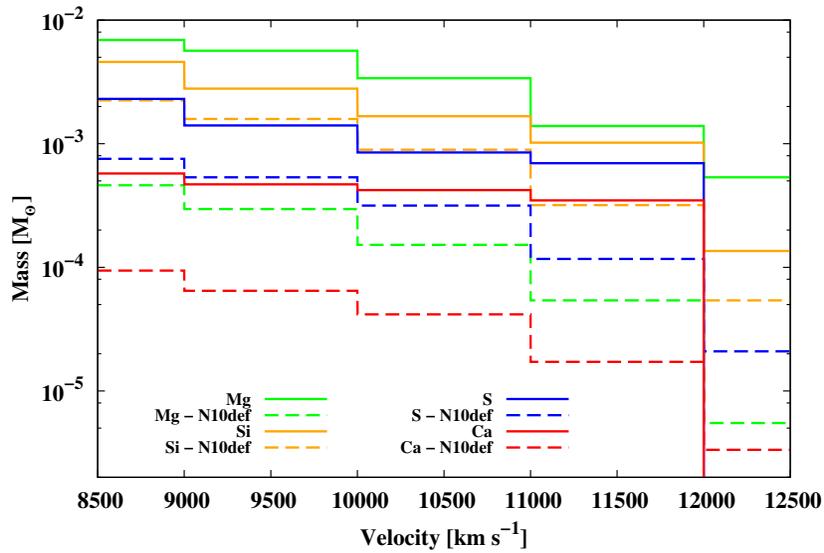}
\caption{\label{fig:fink_abu1} The comparison of IME shell masses we get from the TARDIS fits (solid lines) with those originating from the N10def pure deflagration model (dashed lines).}
\end{figure*}
\begin{figure*}
\centering
\includegraphics[width=11cm]{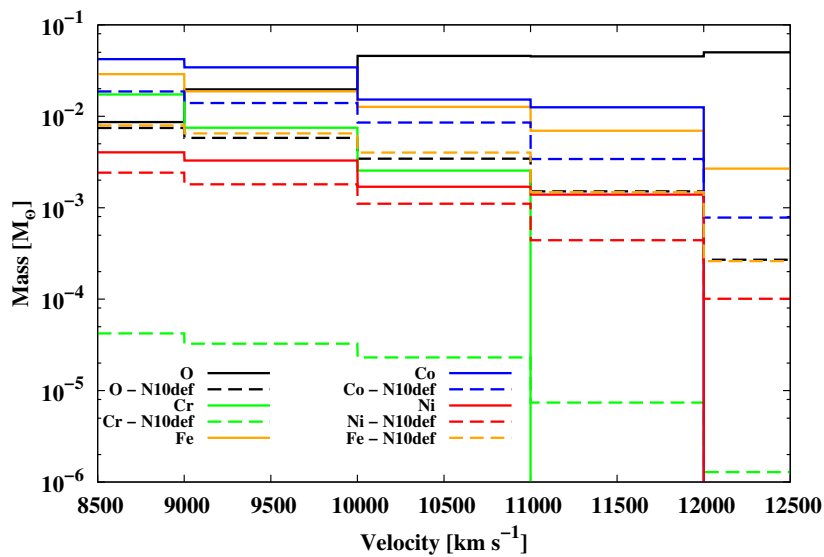}
\caption{\label{fig:fink_abu2} The comparison of IGE and oxygen shell masses we get from the TARDIS fits (solid lines) with those originating from the N10def pure deflagration model (dashed lines) at +7 days regarding to B maximum.}
\end{figure*}

The chemical composition of N10def is also compared to the abundance distribution from the TARDIS fits. Note that we use shell masses rather than mass fractions for this comparison, since the abundance tomography depends not only on the mass fractions but also on the density profile. Since IGEs are present with greater fractions than IMEs, we display the shell masses of the two groups in two different figures (Figs. \ref{fig:fink_abu1} and \ref{fig:fink_abu2}). As can be seen in both figures, the masses from the TARDIS model are higher than those from the deflagration model N10def; may be partly caused by the differences in the density profiles, although, the trends of the two distributions are similar.

Looking at the IMEs, the shell masses of the TARDIS model are higher by nearly one order of magnitude, except in the case of Si, where the difference is only a factor of two. Note that the S and Ca distributions cut off at 12,000 km s$^{-1}$, which does not happen in the deflagration model. At the same time, there is a larger discrepancy in the amount of Mg above 12,000 km s$^{-1}$. Magnesium appears with a constant mass fraction of $\sim$0.01 in the deflagration model, while the TARDIS model shows an increasing mass fraction of magnesium toward the lower velocity regions.
This could explain the lack of the strong Mg\,\begin{small}II\end{small} feature at $\sim$8,800 \r{A} in the deflagration model spectra, which is properly fit by the TARDIS model.

Taking a look at Fig. \ref{fig:fink_abu2}, it can be seen that the distributions show mostly the same trend in the case of Fe, Co and Ni, but the TARDIS model does not contain any cobalt or nickel above 12,000 km s$^{-1}$.
This seems to contradict the pure deflagration scenario, which predicts high mass fractions of Fe-peak elements at all velocities. Note, however, that this dominance of IGEs in the outer layers does not lead to the emergence of extremely strong Fe/Ni/Co lines in the N10def model, see Fig. \ref{fig:fink_spectra} (while the presence of early, excessively strong Fe lines is an argument against uniform abundance TARDIS models, see text and Fig. \ref{fig:spectra} in Sec. \ref{tardisfitting}).

In Fig. \ref{fig:fink_abu2} we also show the mass of chromium which has a mass fraction of $\sim$0.001 in the whole ejecta based on the N10def model. This value is two orders of magnitude less than what we use in \begin{small}TARDIS\end{small} to fit the blue part of the spectra obtained at +7 and +10 days. This discrepancy makes the high amount of chromium in the TARDIS model questionable, because only the flux depression between 3,500 and 3,800 \r{A} makes its presence necessary. Moreover, only the +7 and +10 days spectra cover this spectral region, which could also be fitted using other elements, like cobalt or titanium, based on the literature. The best-fit \begin{small}TARDIS\end{small} model contains a high amount of cobalt originating from the nickel-decay, but it shows only a limited contribution to the associated flux depression. Titanium, on the contrary, is not included in our TARDIS model, since it produces strong absorption features in other parts of the observed spectral range, which decrease the goodness of the fit. Nevertheless, since chromium and titanium are not synthesized independently during the nuclear processes, we cannot exclude the possible presence of Ti in the ejecta of SN~2011ay.

The oxygen masses from the two models are also plotted in Fig. \ref{fig:fink_abu2}. Since we use the oxygen as a ''fill up'' element (see Sec. \ref{elements}), this comparison is less informative. Thus we avoid to take any conclusion about the amount of oxygen.

There are two main findings of our comparative analysis. First, it seems that the pure deflagration models published by \citet{Fink14} predict a faster evolution than shown by the observed light curves of SN~2011ay. This may be caused by the ejecta mass being too low for the deflagration models. Secondly, the comparison of the chemical compositions of the ejecta implied by the N10def and \begin{small}TARDIS\end{small} models, shows some notable differences in the amount and/or distribution of both IMEs and IGEs.

Both findings indicate that explosion scenarios other than pure deflagration should be taken into account to explain the origin of SN~2011ay. The direct comparison of our data with these alternative scenarios, presented in detail in Sec. \ref{introduction}, is beyond the scope of our current paper; the reason for that is the lack of available model spectra or light curves regarding SNe Iax. Nevertheless, we briefly compare our findings with the predictions of the different explosion theories.

To achieve larger ejecta masses, delayed detonation models are promising alternatives. However, we note again that modified pure deflagration models \citep[with ignition conditions and/or density parameters different from those used in][]{Fink14} may also lead to higher ejecta masses. In addition, more exotic scenarios (faint CC events with fall-back or mergers of WDs with compact objects) predict low ejecta and $^{56}$Ni masses in general, which means that they may be viable explanations for very low-luminosity SNe Iax but hardly for the brighter events such as SN~2011ay.

As we mention above, the moderate amount of IGEs (more exactly, the lack of Ni and Co) in the outer layer of our TARDIS model ejecta seem to be an argument against pure deflagration models. Moreover, the decreasing trend of mass fractions of IGEs and of some IMEs (Si, Mg) towards higher velocities as well as the "cut-off" of some other IMEs (Ca, S) above 12,000 km s$^{-1}$ indicate a stratified rather than a fully mixed ejecta, at least in the outer parts (see Figs. \ref{fig:VD}, \ref{fig:fink_abu1}, and \ref{fig:fink_abu2}). \citet{Stritzinger15} obtained similar findings for SN~2012Z, which they tried to explain with a one-dimensional pulsational delayed detonation model of \citet{Hoeflich02}. While their applied PDD model seems to serve as a viable explanation for most of the observed properties of SN~2012Z, there are some predictions that are in conflict with the observed features (and also with those of SN~2011ay), e.g. the presence of a relatively high amount of C in the outer layers. Nevertheless, it should be noted that \citet{Stritzinger15} did not fit models to their photospheric spectra, but only to some nebular ones.

Although they describe the explosions of normal Type Ia SNe, we also mention here the three-dimensional hydrodynamic calculations of \citet{Jordan12a} for ''pulsationally assisted'' GCD models, since this scenario has also been mentioned as a possible alternative for the origin of SNe Iax. From their 3D models they find an ''inverse layering'' of the explosion ejecta: the detonation ashes (IGEs surrounded by IMEs and C+O) are enshrouded by IGE-rich deflagration ashes. This model also seems to be in contradiction with our findings.

\section{Conclusions} \label{conc}
We analyze ten spectra of SN 2011ay obtained between -3 and +19 days
with respect to B-max. Based on the photometric and spectroscopic
measurements, SN~2011ay is categorized as one of the most luminous
members of the Iax subgroup with a peak absolute magnitude of
$M\rmn{_{V}}$ = -18.39$\pm$0.18. The main goal of this study is to map
the distribution of chemical elements in the SN ejecta via abundance
tomography using the one dimensional Monte Carlo radiative transfer
code TARDIS.

With the assumed input parameters ($t\rmn{_{exp}}$, W7 density
profile), the manual fitting ($v\rmn{_{phot}}$, $L\rmn{_{e}}$ and
abundance distribution) of the whole spectral dataset becomes
manageable with \begin{small}TARDIS\end{small}. For calculating the abundance distribution, we
have defined a velocity grid with steps of 1,000 km s$^{-1}$, where
the mass fractions of the elements are fitting parameters in each
cell. The final \begin{small}TARDIS\end{small} model shows good
agreement with all the spectra obtained between -3 and +19 days with
respect to B-band maximum. This model has a velocity-dependent,
stratified abundance-structure, which is in conflict with pure
deflagration models, the favoured models for SNe Iax that predict
well-mixed ejecta. We note, however, that our assumed density profile
for the \begin{small}TARDIS\end{small} fitting differs significantly from the density profiles
of the deflagration models, particularly in the outer layers.



%

Building up the TARDIS model atmosphere, we vary the mass fraction of O, Mg, Si, S, Ca, Cr, Fe, Co, and Ni.
Carbon is absent from
our TARDIS models because we were unable to identify carbon features
unambigously in the observed spectra of SN~2011ay.

We have compared the observed light curves and spectra of SN~2011ay to
the synthetic observables of the pure deflagration models of
\cite{Fink14}. Peak values of the synthetic light curves calculated from N10def and
N20def models seem to be comparable with the optical (BVR) data of
SN~2011ay; however, the shapes of the light curves  can not be consistently fit even if we modify the assumed date of explosion. Carrying out a similar comparative analysis of the
observed and synthetic spectra, one finds only poor agreement at early
epochs. Around maximum light, the N10def model spectra show a decent
fit below $\sim$6,500 \r{A}. This remains true even at the observed
post-maximum phases.

The abundance distribution of the best-fit stratified
\begin{small}TARDIS\end{small} model is also compared to the abundance
profile originating from the N10def model. The distributions of IMEs
differ only slightly from the hydrodynamic calculations; the only
discrepancy appears in the fraction of magnesium. At lower velocities
(< 10,000 km s$^{-1}$), the abundances of IGEs are also similar to the
results from the N10def model. On the other hand, the decreasing
trends of IGE abundances toward the outer regions (above 10,000 km
s$^{-1}$) cannot be seen in the N10def model that predicts a
well-mixed ejecta. This outer region, which seems to be
stratified rather than mixed according to the best-fit
\begin{small}TARDIS\end{small} model, has an impact mainly on the
spectra obtained at the early epochs. Therefore, this difference in
the abundance distribution could explain why the \begin{small}TARDIS\end{small} model provides
a superior fit to SN~2011ay at early epochs, compared to N10def.

%
%
%
As a conclusion, our comparative analysis of
\begin{small}TARDIS\end{small} and pure deflagration hydrodynamic
models shows systematic differences in the composition of the outer
ejecta layers. Given the modelling uncertainties and particular the
difference in the underlying density profiles of the TARDIS and hydro
models, an interpretation of this result is not straightforward.
However, it could indicate that the propagation of the thermonuclear
flame may differ from a pure deflagration scenario in this case.
Nevertheless, it should be highlighted again that SN~2011ay seems to be an extreme member of SNe Iax, considering its relatively high luminosity, expansion velocity and peculiar spectroscopic nature in the nebular phase. Thus, we are reticent to draw any general conclusions regarding the whole class of these objects. In order to do that, it seems to be necessary to examine various members of the Iax class using the method of abundance tomography, which we intend to present in a forthcoming paper.

\section*{Acknowledgments}
B.B. received support from the Campus Mundi Short Study Programme of Tempus Public Foundation.
T.S. is supported by the NKFIH grant PD112325 of the Hungarian National Research, Development and Innovation Office. 
MK acknowledges support from the Klaus
Tschira foundation.
J.M.S. is supported by an NSF Astronomy and Astrophysics Postdoctoral Fellowship under award AST-1302771.
W. K. is supported by ESO Fellowship $\&$ Google Summer of Code.
This work is also supported by the GINOP-2.3.2-15-2016-00033 grant of the Hungarian National Research, Development and Innovation Office.

\end{document}